# Drive Current Boost in Double-Channeled Nanotube Gate all Around Field Effect Transistor

Laixiang Qin[1], Chunlai Li[2], Yiqun Wei[2*], Zhangwei Xu[1], Jin He[1*], Yandong He[1], Yutao Yue[2]

*1.Shenzhen SoC Key Laboratory, PKU-HKUST Shenzhen-Hong Kong Institution, Shenzhen 518057, China*

*2.Shenzhen institute of Peking University, Shenzhen 518057, China*

[*] Electronic email: frankhe@pku.edu.cn; weiyq@ier.org.cn

## Abstract

Gate all around field effect transistor (GAAFET) presents a resurgence ascribed to its enhanced gate electrostatic controllability by virtue of surrounding gate structure in coping with increasingly serious power consumption dissipation and short channel effects (SCE) degradation as the semiconductor technology enters into sub-10nm technology node. Nanotube GAAFET (NT GAAFET) with inner and outer channels surrounding by inner and outer gates proves to be superior than GAAFET in drive current ($I_{on}$) augmentation and SCEs inhibition attributed to enhanced gate electrostatic integrity, holding promise to expand the Moore's law Roadmap further beyond. Herein, we demonstrate an exotic doubled-channeled NT GAAFET (DC NT GAAFET) structure with $I_{on}$ boost in comparison with NT GAAFET and NW GAAFET with the same footprint. $I_{on}$ gains of 64.8% and 1.7 times have been obtained in DC NT GAAFET in compared with NT GAAFET and NW GAAFET. $I_{off}$ of DC NT GAAFET degrades by 61.8% than that of NT GAAFET, SS is almost comparable in two kinds of device structures, whereas $I_{on}/I_{off}$ ratio in DC NT GAAFET still gains subtly, by 2.4%, than NT GAAFET thanks to the substantial $I_{on}$ aggrandizement, indicating the sustained superior gate electrostatic controllability in DC NT GAAFET with regarding to NT GAAFET regardless of additional channel incorporated. On the other side, both DC NT GAAFET and NT GAAFET exhibit superior device performance than NW GAAFET in terms of high operation speed and better electrostatic controllability manifested by suppressed SCEs.

## Introduction

Recent years have witnessed a sustained transistor dimensions downscaling to meet quests for high speed, low cost, and low power consumption logic circuits according to Moore's law[1-3], satisfying the demands of emerging internet of thing, big data analytics, intelligent terminals, ultra-large scale computing, and health monitoring, serving the variability and feasibility of today's life[4-5]. Whereas, as semiconductor technology enters into sub-10nm technology node, power consumption dissipation and short channel effects (SCEs) degradation pose substantial challenges for transistor to further scale down. SCEs, composing of drain induced barrier lower (DIBL), subthreshold swing (SS), subthreshold voltage ($V_t$) roll off, leakage current increase and so forth, have long been plaguing transistor by degrading device performance[6-9]. Fortunately, these issues can be circumvented once the channel length is more than 3 to 5 times larger than screen length, $\lambda$, according to recent research study[10-12]. $\lambda$ represents the bending areas between the source to channel and channel to drain in the band diagram, which is expressed as,

$$\lambda = \sqrt{\frac{\varepsilon_{ch}}{N\varepsilon_{ox}} t_{ox} t_{ch}} \qquad (1)$$

$\varepsilon_{ch}$ and $\varepsilon_{ox}$ denote permittivity of channel and oxide materials, $t_{ox}$ and $t_{ch}$ symbolize the dielectric and channel thickness, N is on behalf of gate numbers. To obtain an enhanced SCEs inhibition ability, a smaller $\lambda$ is preferred. Aside from aggrandizing $\varepsilon_{ox}$ with high k dielectric, dwindling $\varepsilon_{ch}$, $t_{ox}$ and $t_{ch}$ as incumbent technology prevalently used, another effective avenue of diminishing $\lambda$ is to increase N according to expression (1), which is thoroughly verified by Fin Field effect transistor (FinFET) with three sides of the channel surrounded by gates exhibiting improved immunity toward SCEs and gate all around field effect transistor (GAAFET) coming under the spotlight deriving from being considered with ultimate gate controllability[13-16]. Motivated by the achievements obtained, another exotic device structure of NT GAAFET with inner and outer channels surrounded by inner and outer gates have been brought about to further enhance the gate to channel coupling effect, intriguing results have been obtained both in aspects of enhanced gate electrostatic integrity and superior scalability[17-21]. NT GAAFET is asserted to possess superior drive current in compared with GAAFET with 50 layers stacked channels of the same footprint. NT GAAFET is proposed to be able to sustain its superior gate electrostatic integrity as long as its NT thickness is equal to or less than half of the gate length ($L_g$)[17].

Inspired by these points, we come up with an exotic double-channeled NT GAAFET (DC NT GAAFET) structure to make full play of the inner vacant space of NT GAAFET to obtain a high drive current without increasing the footprint. The device performances of a NT GAAFET and NW GAAFET of same footprint have been compared with that of DC NT GAAFET with the mean of Technology Computer Aided Design (TCAD) simulation study and $I_{on}$ gains of 64.8% and 1.7 times have been obtained in DC NT GAAFET in compared with NT GAAFET and NW GAAFET. $I_{off}$ of DC NT GAAFET degrade by 61.8% than that of NT GAAFET, which is originated from cumulative channels and slightly decreased threshold voltage ($V_t$), from 0.120 to 0.124V, as is well known that $I_{off}$ increases exponentially with $V_t$ descent following the power law of $I_{off} \propto \exp[-V_t/(KT/q)]$; SS is comparable in both cases, indicating comparable gate electrostatic controllability in DC NT GAAFET and NT GAAFET device structures; $I_{on}/I_{off}$ ratio in DC NT GAAFET still exhibits subtle augmentation, by 2.4%, than NT GAAFET thanks to numerous $I_{on}$ gain brought about by the additional channel. Compared with NW GAAFET with same footprint, both DC NT GAAFET and NT GAAFET exhibit higher operation speed and superior electrostatic controllability manifested by boosted $I_{on}$ and inhibited SCEs.

**Device Structure and Simulation**

The 3-D structures of DC NT GAAFET, NT GAAFET and NW GAAFET and their corresponding cross profiles perpendicular to (y-cuts) and along the channel directions (y axis) simulated in this study are shown in Fig.1. The gate lengths selected for all three GAAFET structures are 50nm, the source/drain lengths are 25nm. Nanotube (NT) thickness of DC NT GAAFET and NT GAAFET is 8nm, the outermost NT radius is 40nm, nanowire (NW) radius is 40nm to make sure that the footprint of NW GAAFET is the same with that of DC NT GAAFET and NT GAAFET. Source/drain have doping concentrations of $1\times10^{20}$/cm$^3$ with arsenic. The channels are doped with boron with concentrations of $1\times10^{16}$/cm$^3$. The high k oxide is HfO$_2$, with a thickness of 1nm. The models used are drift diffusion, high field saturation velocity, inversion and accumulation layer mobility containing doping degradation and Coulomb impurity scattering, normal electric field scattering, Shockley-Read-Hall (SRH) Generation Recombination, Auger Recombination, Bandgap Narrowing models and so forth. In addition, the quantization gradient density model is adopted to incorporate the quantum correction ascribed to carrier confinement effects originating from nanometer-scaled device dimensions.

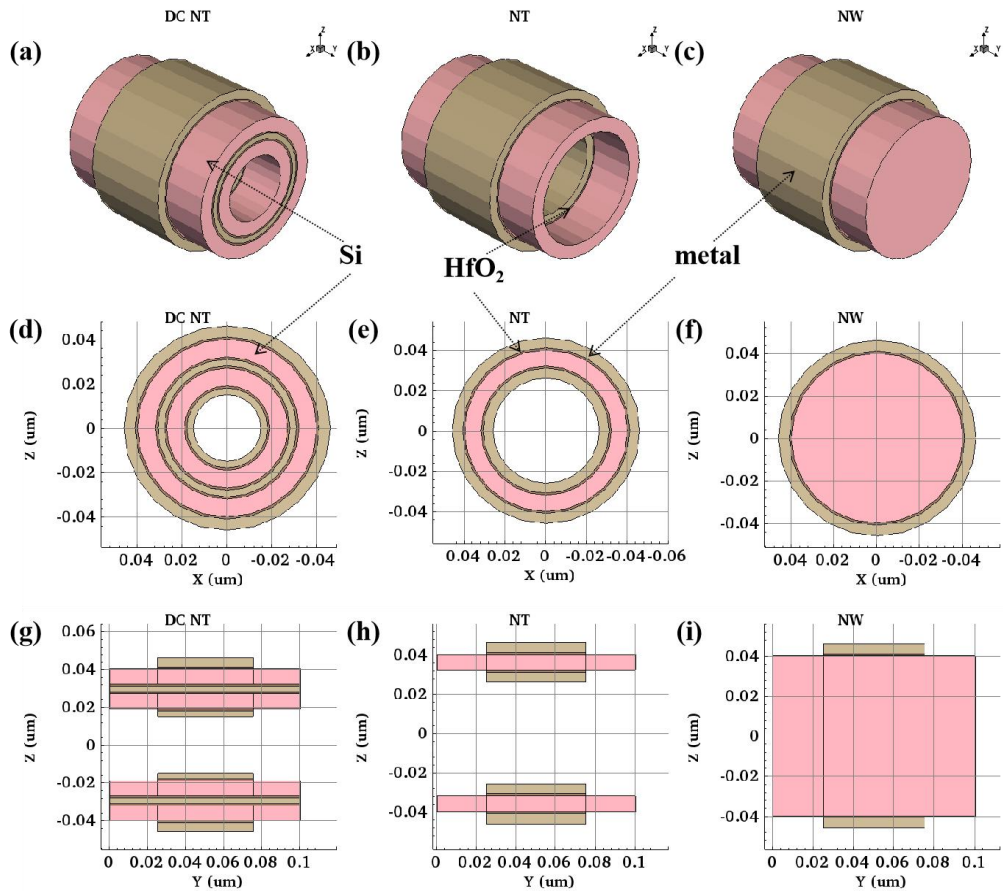

**Fig. 1** 3D structure profiles of (a) DC NT GAAFET, (b) NT GAAFET and (c) NW GAAFET; and corresponding y-cut profiles perpendicular to the channels (d), (e), (f); and x-cut profiles along the channels (g), (h), (i).

**Results and Discussions**

The transfer and output characteristics of DC NT GAAFET (the red lines decorated with circles), NT GAAFET (the blue lines decorated with diamonds) and NW GAAFET (the dark lines adorned with crosses) are demonstrated in Fig.2. Colossal $I_{on}$ boost can be obtained in DC NT GAAFET in compared with NT GAAFET (67.8% more gain) and NW GAAFET (1.7 times larger) thanks to the introduction of the additional channel, whereas the additional channel incorporation also results in degraded $I_{off}$ in DC NT GAAFET by 61.8% than that of NT GAAFET. Consequently, $I_{on}/I_{off}$ ratio in DC NT GAAFET still exhibits subtle gain by 2.4% thanks to the $I_{on}$ gaining to a more tremendous extent than $I_{off}$ degradation. SS is comparable in DC NT GAAFET and NT GAAFET, only decrease from 61.1mV/dec in DC NT GAAFET to 60.9mV/dec in NT GAAFET, indicating the condign gate electrostatic controllability in both devices. As for the NW GAAFET of the same footprint, not only $I_{on}$ descends enormously than both DC NT GAAFET and NT GAAFET, but also SS and $I_{off}$ degrade 5 times and more than one order than those of DC NT GAAFET and NT GAAFET, corroborating the superior gate electrostatic controllability of both DC NT

GAAFET and NT GAAFET than NW GAAFET attributed to thinner channel thickness and double-gated surrounded channel structures.

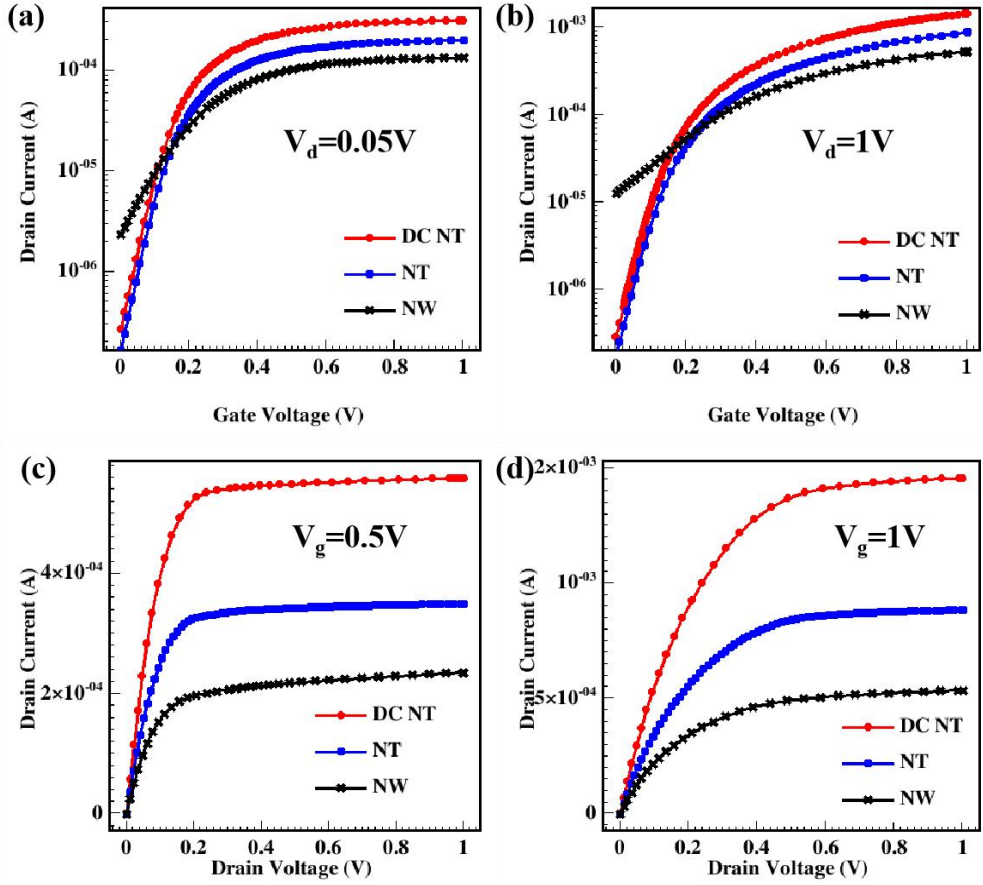

**Fig.2** Transfer characteristics of DC NT GAAFET (the red lines), NT GAAFET (the blue lines) and NW GAAFET (the dark lines) under $V_{ds}$ of 0.05V (a) and 1V (b) respectively; Output characteristics of DC NT GAAFET (the red lines), NT GAAFET (the blue lines) and NW GAAFET (the dark lines) under $V_{gs}$ of 0.5V (c) and 1V (d) respectively.

To elucidate the mechanism of $I_{on}$ boost in DC NT GAAFET than both the NT GAAFET and NW GAAFET, electron current density distribution profiles perpendicular to and along the channel directions are depicted in Fig.3. The impact of the additional channel introduced in the inner vacant space of the core region of the NT GAAFET, ie, DC NT GAAFET structure, on the electron current/electron density distributions in two devices cases is subtle, indicating the comparative gate electrostatic control in DC NT GAAFET and NT GAAFET, as further manifested by comparable SS and $I_{on}/I_{off}$ ratio. In both the DC NT and NT GAAFET conditions, the double gated structure fuel partial or even total depletion in the channel region depending on the NT thickness and gate stack structures, giving rise to larger concentration of minority carriers in proximity to the channel in comparison with NW GAAFET, which are hypothesized to be capable of contributing to $I_{on}$ surge in contrast to minority carriers that penetrating far into the channel as in the case of NW GAAFET, as displayed in Fig.3 (g), and (h). For another, partial or total depletion of the channels in DC NT GAAFET and NT GAAFET brought about by double gated structures also contribute to electron mobility, μ, increasing immensely attributable to less scattering both arising from surface and traps lying in the interfaces between semiconductor channel and insulator, as shown in Fig.3 (i). These two points as well as widened effective channel width, $W_{eff}$, account for the $I_{on}$ surge in DC NT GAAFET and NT GAAFET as opposed to NW GAAFET according to the expression of $I_{on}$ depicted in (2). The $I_{on}$ boost in DC NT GAAFET with regarding to NT GAAFET mainly ascribes to $W_{eff}$ increment, as well elucidated by equation (2). More details can be referred to our former published paper[21], where the mechanism of $I_{on}$ boost of doubled gated NT GAAFET in comparison with NW GAAFET has been intensively discussed.

$$I_{DSAT} = \frac{W}{2L}\mu C_{OX}(V_{GS} - V_{TH})^2 \qquad (2)$$

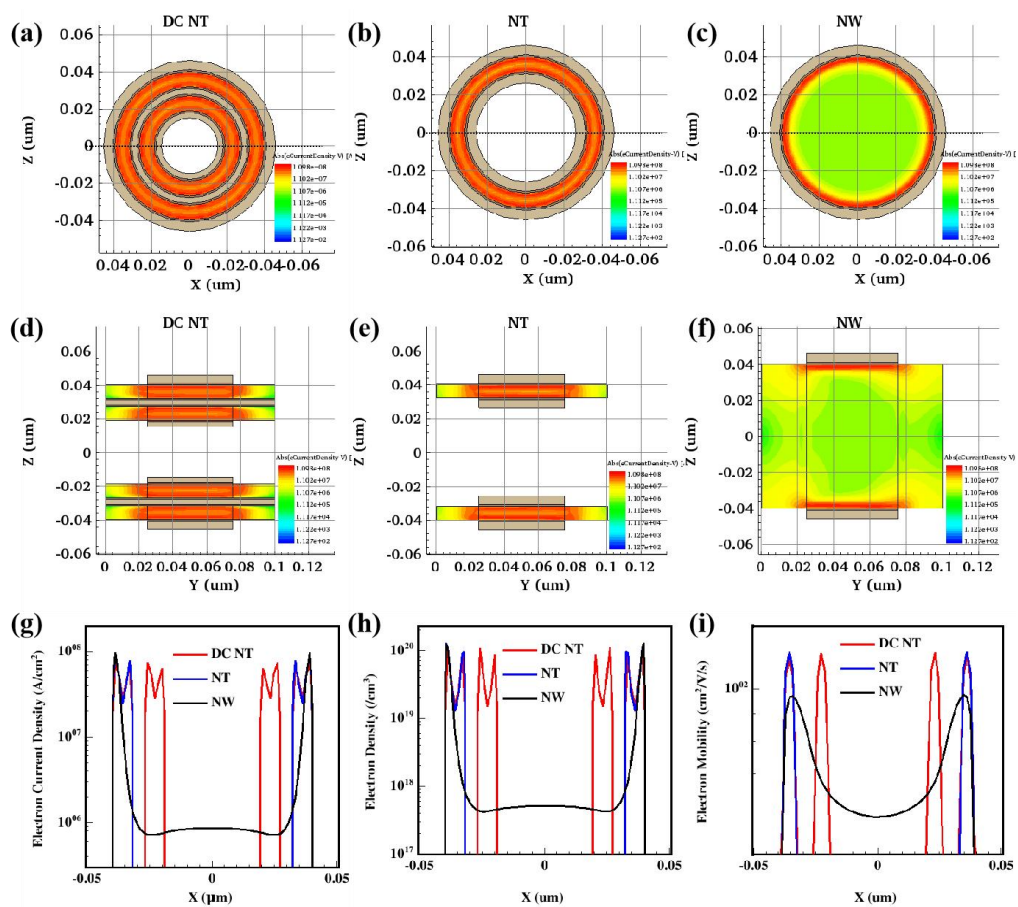

**Fig.3** Electron current density profiles of DC NT GAAFET, NT GAAFET, NW GAAFET perpendicular to (a), (b), (c) and along (d), (e), (f) the channel directions; electron current density (g), electron density (h), and electron mobility distribution profiles along the cut line as shown in (a), (b), (c).

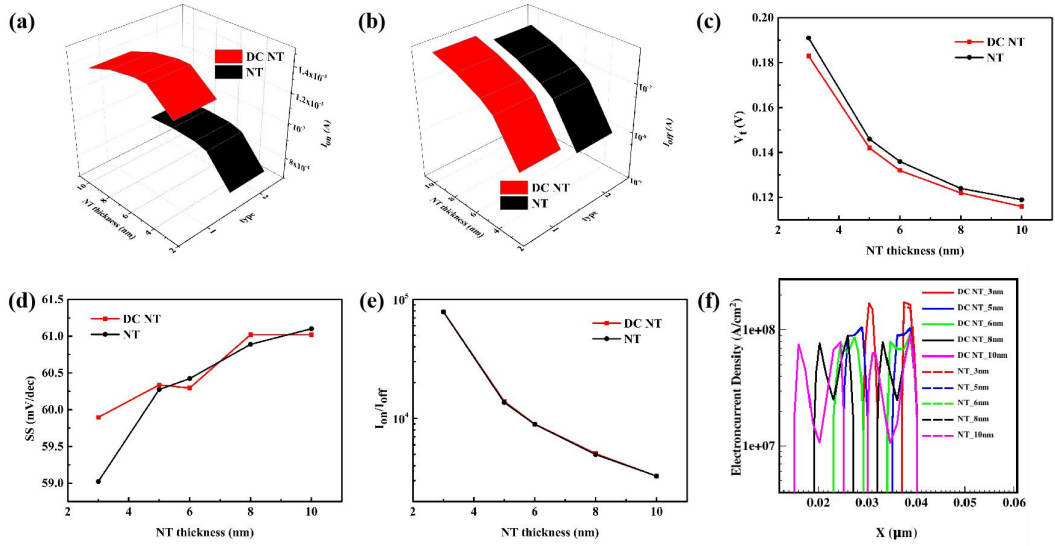

**Fig.4** Performance dependence comparisons on NT thickness in DC NT GAAFET and NT GAAFET. (a) $I_{on}$; (b) $I_{off}$; (c) $V_t$; (d) SS; (e) $I_{on}/I_{off}$ ratio vs NT thickness trends of DC NT GAAFET in comparison with NT GAAFET respectively; (f) Electron current density distribution profiles of DC NT GAAFET (the solid lines) and NT GAAFET (the dashed lines) with different NT thickness along the cutlines as shown in Fig3. (a).

Fig.4 summarizes the device performances and SCEs dependence on NT thickness in DC NT GAAFET and NT GAAFET with NT thickness varying from 3 to 10nm. DC NT GAAFET demonstrates dramatic $I_{on}$ gains than NT GAAFET in all NT thickness simulated thanks to the additional channel, as depicted in Fig.4 (a), while on the other hand, the additional channel introduction also leads to $I_{off}$ degradation in DC NT GAAFET. Besides, relatively smaller $V_t$ in DC NT GAAFET also results in larger $I_{off}$, as shown in Fig.4 (c). As for SS, both DC NT GAAFET and NT GAAFET show a SS improvement with NT thickness scaling down, which can be well accounted for by better gate electrostatic controllability arising from thinner channel thickness. When NT channel thickness scales down to 3nm, SS of both DC NT GAAFET and NT GAAFET drop down below 60mV/dec, which are surmised to be attributed to tunneling current participating in current drive at such small channel thickness. And for another, SS of DC NT GAAFET is comparable with that of NT GAAFET at larger channel thickness, indicating condign gate electrostatic controllability in this channel thickness regime; nonetheless, SS in DC NT GAAFET differs from that of NT GAAFET when channel thickness scales down below 5nm, we attributed this phemomena at thinner channel thickness to enhanced quantum interaction effect in thinner channels as well as between adjacent channels. $I_{on}/I_{off}$ ratio in both DC NT GAAFET and NT GAAFET structures increase monotonically with NT thickness scaling down ascribing to strengthened gate electrostatic control. For another, $I_{on}/I_{off}$ ratio of both DC NT GAAFET and NT GAAFET stay comparable, on account that though $I_{on}$ and $I_{off}$ of DC NT GAAFET increase concurrently attributable to the additional channel, and $I_{on}$ gain and $I_{off}$ degradation in compared with those of NT

GAAFET are almost of the same proportions, thus almost coincident $I_{on}/I_{off}$ ratio in two device structures are exhibited vs NT thickness trends as demonstrated in Fig.4 (e), indicating superior gate electrostatic controllability sustained in DC NT GAAFET device structure despite additional channel incorporated, posing encouraging promise for further enhancing $I_{on}$ without SCEs degraded. Fig.4 (f) demonstrates the electron current distribution profiles of DC NT GAAFET and NT GAAFET under different NT thicknesses along the right half cutlines as shown in Fig.3(a). Except for an additional inner channel electron current distribution profiles, the electron current distribution profiles in outer channels of DC NT GAAFET are almost coincident with those of NT GAAFET at the corresponding space at almost all NT thickness, only slight noncoincidences occur at thinner NT thickness, which is supposed to be attributed to enhanced quantum interaction effect occurring between adjacent channels, indicating the comparable gate electrostatic control of every single channel in DC NT GAAFET with that of NT GAAFET, casting a new light on building further $I_{on}$ boosting device structures with more additional channels integrated and even various kinds of logic circuits by making full play of the inner vacant spacer of NT GAAFET with neither gate electrostatic control weakened nor footprint augmented.

**Conclusion**

In conclusion, we demonstrate a fascinating DC NT GAAFET device structure, which holds the promise of boosting $I_{on}$ prodigiously without degrading SCE obviously than NT GAAFET and NW GAAFET device structures. $I_{on}$ gains of 64.8% and 1.7 times have been obtained in DC NT GAAFET in comparison to NT GAAFET and NW GAAFET. $I_{off}$ and SS of DC NT GAAFET degrade slightly than those of NT GAAFET, whereas $I_{on}/I_{off}$ ratio still gains a little due to the substantial $I_{on}$ boost of DC NT GAAFET. Both DC NT GAAFET and NT GAAFET exhibit superior device performance than NW GAAFET in regard to high speed and better electrostatic control manifested by suppressed SCEs. The exotic DC NT GAAFET structure holds the promise of building further $I_{on}$ boosting device structures or simple logic circuits, take complementary FET(CFET) as an example, with neither gate electrostatic control damaged nor footprint augmented.

**Acknowledgment**

This work is supported by these funds of 2021Szvup002, JCYJ20200109144612399, JCYJ20200109144601715, JCYJ20210324115812036, JCYJ20220818103408017, JCYJ20220818103408018, IERF202002, IERF202105, IERF202206, JCYJ20220818103408018.